\documentclass[12pt]{article}

\usepackage{hyperref}

\begin{document}

\title{Computer algebra in gravity\footnote{Updated version of an
article to appear in: {\em Computer Algebra
Handbook}. J.~Grabmeier, E.~Kaltofen, V.~Weispfennig, editors (Springer,
Berlin 2001/2002).}}
\author{Christian Heinicke\footnote{Email: chh@thp.uni-koeln.de}\\
       Friedrich W.\ Hehl\footnote{Email: hehl@thp.uni-koeln.de}\\
Institute for Theoretical Physics\\
University of Cologne\\
50923 K\"oln, Germany}

\maketitle

\begin{abstract}
We survey the application of computer algebra in the
context of gravitational theories. After some general remarks, we show of 
how to 
check the second Bianchi-identity by means of the Reduce package Excalc.
Subsequently we list some computer algebra systems and packages relevant
to applications in gravitational physics. We conclude by presenting a couple 
of typical examples.  

\bigskip

\bigskip

\centerline{\bf Contents}

\noindent
1. \hyperlink{sec1}{Introduction}\\
2. \hyperlink{sec2}{Riemannian curvature in tensor and exterior calculus}\\
3. \hyperlink{sec3}{Abstract and component CA systems}\\
4. \hyperlink{sec4}{General versus special purpose systems}\\
5. \hyperlink{sec5}{Applications}\\
6. \hyperlink{ref}{References}
\end{abstract}

\newpage


\section{Introduction}


\hypertarget{sec1}{Einstein's} gravitational theory,  {\em general relativity} (GR),
is the valid theory for describing gravitational effects.
In the search for
making GR compatible with quantum theory and/or unifying it with the other
interactions of nature (strong, electro-weak, superweak,\dots), different
schemes have been developed, like the gauge approach to gravity, including
supergravity and metric-affine gravity, higher-dimensional Kaluza-Klein type
models, string models, but also more conventional Hamiltonian (canonical) or
Feynman quantization schemes or, more far-fetched, models based, e.g., on
noncommutative spacetime geometries.

The computer algebra (CA) programs applied in GR can be and partially have 
been extended to these more general frameworks. Still, it is probably  true 
that most CA programs in gravity are applied in the context of GR followed by
those for evaluating gravity-based Feynman integrals and for executing
computations in the framework of 
gauge models encompassing non-Riemannian spacetimes. In this
note we mainly concentrate on GR and will give a couple of examples.

Detailed overviews of computer algebra in GR
are given, e.g., by Brans\,\,\cite{brans}, Hartley \cite{hartley}, 
Lake \cite{lake}, and 
MacCallum \cite{mac1}, \cite{mac2}.


\section{Riemannian curvature in tensor and exterior calculus}


\hypertarget{sec2}{In} GR and gravity, computer algebra was used as soon as it
became available. The reason for this is that for solving standard problems
it is required to manipulate a large number of terms and equations.
We will clarify this by an example. A generic problem in gravity is to 
calculate the Ricci tensor  from a given metric. 
A general form of a spacetime metric
$g$ in four dimensions is given by 10 independent functions $g_{ij}=g_{ji}$ 
of the coordinates $(x^0,x^1,x^2,x^3)$:
\begin{equation}\label{metric}
g=\sum_{i,j=0}^{3} g_{ij}(x^0,x^1,x^2,x^3) \, 
                 {\rm d}x^i \otimes {\rm d}x^j\,\,.
\end{equation}
The so--called Christoffel symbols are determined 
from the functions $g_{ij}$ by means of the following equations:
\begin{equation}\label{chris}
 \Gamma_{ij}{}^k =  \frac{1}{2} \sum_{m=0}^3 g^{km} \, \left(
         \frac{\partial\,g_{jm}}{\partial x^i}
        +\frac{\partial\,g_{mi}}{\partial x^j}
        -\frac{\partial\,g_{ij}}{\partial x^m}
        \right) = \Gamma_{ji}{}^k\,\,.
\end{equation}
Here $g^{km}$ denotes the matrix reciprocal to $g_{ij}$.
Since $i,j,k$ are running from $0$ to $3$, the $\Gamma_{ij}{}^k$ represent 64
functions. Because of the symmetry in $i,j$, only 40 are independent.
In the conventions of Schouten \cite{schouten},
the Riemannian curvature tensor is derived from the Christoffel symbols 
in the following way:
\begin{equation}\label{curv}
R_{ijk}{}^{l} =  \frac{\partial \,\Gamma_{jk}{}^l}{\partial x^i}
               -  \frac{\partial \,\Gamma_{ik}{}^l}{\partial x^j}
               + \sum_{m=0}^3 \left( \Gamma _{im}{}^l \,\Gamma _{jk}{}^m
                                    -\Gamma _{jm}{}^l\, \Gamma
_{ik}{}^m\right)\,\,.
\end{equation}
Eventually we find the Ricci tensor as
\begin{equation}\label{ricci}
{\rm Ric}_{jk} = \sum_{i=0}^3 R_{ijk}{}^i.
\end{equation}
Now one can easily estimate that the number of terms in each of the
components ${\rm Ric}_{ij}$ may be very large. In \cite{lake2} 
it is shown that in the general case
this number is in the order of  10 000 for each of the components.
Thus, only in simple cases can these calculations be done conveniently 
by hand.

It would be most desirable to have a computer algebra system which
allows to enter mathematical expressions in an analogous way as one would 
write them down on paper. That is, defining objects with abstract properties,
doing calculations, and assigning explicit values to these
objects should be possible in a natural way. Such kind of systems already
exist. We illustrate this by an example.

In terms of Cartan's calculus of exterior differential forms, Eq.(\ref{curv}) 
can be displayed very compactly as 
\begin{equation}
  \label{cartan}
  R_\alpha {}^\beta = {\rm d} \Gamma _\alpha {}^\beta
                    -  \Gamma _\alpha {}^\gamma \wedge \Gamma _\gamma
{}^\beta \,\,.
\end{equation}
Here, $R_\alpha {}^\beta$ is the curvature 2-form and $\Gamma _\alpha{}^\beta$
the connection 1-form. The summation convention is assumed, 
i.e. summation is understood over the repeated index $\gamma$.
Let us check the \emph{Bianchi identity}
\begin{equation}\label{bianchi}
{\rm d} R_{\alpha} {}^{\beta} 
- \Gamma_{\alpha}{}^{\gamma} \wedge R_{\gamma}{}^{\beta} 
+ \Gamma_{\gamma}{}^{\beta}  \wedge R_{\alpha}{}^{\gamma} 
=0
\end{equation}
on the computer. This can be deduced from  Eq.(\ref{curv}) and the 
properties of the exterior product and the exterior derivative. 
In Excalc, a Reduce package for exterior calculus, we first have to
declare that all indices run from $0$ to $3$.
Then we declare  $\Gamma _\alpha{}^\beta$ to be a 1-form and  
$R_{\alpha} {}^{\beta}$ a 2-form.
\begin{verbatim}
indexrange 0,1,2,3;
pform gamma(a,b) = 1 , curv(a,b) = 2;
\end{verbatim}
Eq.(\ref{cartan}) and the left hand side of Eq.(\ref{bianchi}) 
can almost literally be translated,
\begin{verbatim}
curv(-a,b) := d gamma(-a,b) - gamma(-a,c) ^ gamma(-c,b);
\end{verbatim}
and
\begin{verbatim}
d curv(-a,b) - gamma(-a,c) ^ curv(-c,b) + gamma(-c,b) ^ curv(-a,c) ;
\end{verbatim}
The negative sign in front of an index indicates that it is a subscript
whereas a superscript is denoted by a positive (or no) sign. 
The last command yields zero, i.e. the Bianchi identity is confirmed.
In \cite[p.234]{pc} it is shown how to handle this in the 
Mathematica package MathTensor.


\section{Abstract and component CA systems}


\hypertarget{sec3}{Packages} (or systems) capable of performing symbolic 
calculations, as with the Bianchi identity above,
are called \emph{abstract} or \emph{indicial} calculus systems~\cite{hartley}. 
They are necessary if one wants to investigate general
properties of objects. 
So called \emph{component} calculus systems are designed to calculate
the components of unknown quantities from known ones. 
In our example, it was not necessary to introduce a basis or a metric
in order to
define $\Gamma_{\alpha}{}^{\beta}$ and $R_{\alpha}{}^{\beta}$. 
In a typical component system, one first would have to enter the components of
a metric. Then, by means of build--in routines,
the components of the connection and curvature  could be calculated.
Both, abstract as well as component systems, allow to define 
new objects by means of various mathematical operations like products or
derivatives, {e.g.}. The difference here is that in the case of a 
component system  it is always necessary to assume a specific basis and/or 
metric. In turn,
there are abstract calculus systems which do not support computations of
explicit values of components. 
Some packages, like MathTensor or Excalc, for instance, allow both, abstract
and component calculations (see Table 1).
In our Excalc example, we could assign explicit expressions
to the components of $\Gamma_{\alpha}{}^{\beta}$. Then we would find the
components of $R_{\alpha}{}^{\beta}$ by calling \verb|curv(-a,b)|.


\section{General versus special purpose systems}


\hypertarget{sec4}{Today}
 it seems that most people use relativity packages of  \emph{general
purpose systems} like Maple, Mathematica, Reduce, Macsyma and/or Derive.
These programs offer very user-friendly front-ends and, moreover, a wealth of
useful facilities, like simplification routines, programs for solving
algebraic or differential equations exactly or numerically, Tex and Fortran
interfaces, etc.. 

\emph{Special purpose systems} are, as the name suggests, specialized to
handle only a specific class of problems. Therefore the set of instructions
is usually very limited and  the 
programming of these systems requires normally much
more effort than is the case with general purpose systems.  
However, these systems are rather compact, very fast, and may sometimes be the 
only available facility to solve a problem. In the case of calculating
Feynman diagrams, e.g., special purpose systems like Schoonship or Form
are often used, but there are also packages available for Mathematica and
Reduce. A fairly widely used special purpose system for tensor calculus 
and general relativity is Sheep. 
For general relativity formulated in terms of orthonormal frames, the
special purpose system Ortocartan has been recently
updated~\cite{krasinski}.\\
\begin{center}
\footnotesize
\begin{tabular}{|l|l|l|} 
\hline
System & Component & Abstract \\
\hline
Macsyma \cite{macsyma}
                                & CTensor & ATENSOR\\
                      &         & ITENSOR\\
                      &         & CARTAN\\   
\hline
Maple\cite{maple}
                   & tensor & difforms\\
                   & cartan & forms\\
                   & NPspinor & \\
                   & debever & \\
                   & oframe & \\
                   & GRTensorII \cite{grtensor}&\\
                   & Riemann \cite{riemann} & \\
\hline
Mathematica\cite{mathematica}
&  Cartan \cite{sol} & EinS \cite{EinS} \\
                              &   TTC \cite{TTC}
                           & Ricci \cite{ricci}\\
                       & MathTensor \cite{pc}, \cite{mathtensor}& MathTensor\\
                       &                           & DifferentialForms\\
\hline
Reduce \cite{reduce}  
& EXCALC Chap.   & EXCALC\\
                      & REDTEN\cite{redten}  & RICCIR \cite{riccir}\\
                      & GRG      & GRGlib \cite{DermottCA}\\
                      & GRG$_{\it EC}$ & \\
\hline
Sheep \cite{sheep}, \cite{mac1}, \cite{sheep2}  & CORD  & STENSOR\\
                  & FRAME &  \\
\hline
\end{tabular}\normalsize
\end{center}
\pagebreak


\section{Applications}


\hypertarget{sec5}{Today} many authors use computer algebra in order
to obtain or confirm their results
\emph{without} mentioning it explicitly. 
In the following we present some articles
which explicitly illustrate the applications of computer algebra in gravity.

It was already stated that quantities like the \emph{Ricci tensor} can reach 
an enormous size. Thus, for most applications (classification, numerics), these 
objects have to be put on the computer. The use of computer algebra has
the advantage that there is no need to enter  very large
expressions for the curvature, e.g., but a comparably small input program
which calculates these. Moreover, one can use the possibility to transmit
programs or results by means of computer networks, 
cf. the closing remark in \cite{egh}. 

A standard application of computer algebra in GR is the \emph{classification
of exact solutions}. 
These are necessarily found in special coordinates.
However, it is desirable to characterize the
corresponding spacetimes in a coordinate independent way. This includes
determination of Petrov and Segre types, calculations of curvature
invariants (the trace of the Ricci tensor of Eq.(\ref{ricci}), for instance), 
the maximal isotropy group, etc..
The appropriate algorithms involve an enormous amount of calculational work. 
Programs for the widely used Petrov classification
are available for most computer algebra systems.
By means of the Sheep-package Classi \cite{mac1}, it was possible to
create a searchable  online-databank which includes nearly 200  
exact solutions and their properties and which is still growing \cite{jimsk}.

To find out whether two solutions which look different are not just the same
solution in different coordinates, one has to solve the so-called 
\emph{equivalence problem}. 
This involves differentiation of the 
curvature tensor up to the seventh order. In \cite{mac1} appropriate
programs are presented for Sheep.

Computer algebra is also very useful for finding \emph{new solutions} of 
the Einstein equation. For vacuum, it reads ${\rm Ric}_{ij}=0$.
As we can recognize from Eq.(4) together with Eqs.(2) and (3), it
represents a system of ten second order quasi-linear 
partial differential equations for the $g_{ij}$
which are obviously very difficult to solve.
A simple example is given in \cite{vulcanov} where the Schwarzschild
solution is derived from a spherical symmetric line element by using the
Reduce package Excalc.\index{Excalc}
In \cite{ter} it is illustrated how to use the Reduce-based system 
GRG$_{\rm EC}$ for searching for
solutions of the Einstein-Maxwell equations.
In \cite{wolf} the Reduce package Classym was used to derive the
Killing vector and Killing tensor equations and their
integrability conditions from a general form of a metric.
Subsequently the Reduce package Crack has been used for solving these
equations.

In \cite{smh} it is shown how to construct solutions of \emph{metric-affine}
gravity from solutions of the Einstein-Maxwell equations under heavy use of
the computer algebra system Reduce.

A further application is the derivation of a field equation from an action
principle by means  of \emph{variational calculations}.
A simple example is given in \cite[p.303]{pc}
where the Einstein vacuum equation is rederived. In \cite{tsantilis},
the gravitational field equation of a unified field theory of Paw{\l}owski
and R\c{a}czka were derived (and corrected) from the corresponding
Lagrangian by means of MathTensor.

In \cite{deven} an example is given of how to use the special purpose system
Form to calculate \emph{Feynman diagrams}
in the context of quantum gravity.

Another feature of many computer algebra systems is a \emph{Fortran interface}
which converts equations into Fortran readable form. This helps to develop
programs for numeric calculations, as illustrated in \cite{seidel}. 

In \cite{davies} it is outlined of how to apply the computer algebra system
GRTensorII to second-order black hole perturbations.


\begin{thebibliography}{10}
\hypertarget{ref}{}

\bibitem{TTC}
A.~Balfag\'{o}n, P.~Castellv\'{i}  and X.~Ja\'{e}n:
{\em TTC: Symbolic Tensor Calculus with Index.} 
Available at: \href{http://baldufa.upc.es/ttc/}{\mbox{\tt http://baldufa.upc.es/ttc/}}

\bibitem{seidel}
S.~R. Brandt and E.~Seidel:
{\em The evolution of distorted rotating black holes I: Methods and tests.}
Phys.Rev. {\bf D 52} 856--69 (1995).\hfill{}\\
Also available at: \href{http://arXiv.org/abs/gr-qc/9412072}{\tt
http://arXiv.org/abs/gr-qc/9412072}

\bibitem{brans}
C.~H. Brans: 
{\em Computer algebra and general relativity},
in J.~Fleischer, J.~Grabmeier, F.~W. Hehl, and W.~K\"uchlin, editors,
  {\em Computer Algebra in Science and Engineering}, pages 183--195. World
  Scientific, Singapore, 1995.


\bibitem{mathtensor}
S.~Christensen:
{\em MathTensor online documentation}.\\
Available at: \href{http://smc.vnet.net/MathSolutions.html}{
\mbox{\tt http://smc.vnet.net/MathSolutions.html}}

\bibitem{web1}
{\em Computer algebra information network}.\hfill{}\\
Available at: \href{http://www.can.nl/}{\mbox{\tt http://www.can.nl/}}

\bibitem{davies}
G.~Davies:
{\em Second-order black hole perturbations: A computer algebra approach,
{\rm I} -- The Schwarzschild spacetime} (1998).\hfill{}\\
Available at: \href{http://xxx.lanl.gov/abs/gr-qc/9810056}{
\mbox{\tt http://xxx.lanl.gov/abs/gr-qc/9810056}}

\bibitem{egh}
F.~J.\ Ernst, A.~D.\ Garcia, and I.~Hauser,
Journal Math.\ Phys. {\bf 28} 2155--2161 (1987).

\bibitem{vulcanov}
F.~Ghergu and D.~Vulcanov.
{\em Use of computer facilities in teaching general relativity} (1998).
\hfill{}\\
Available at: \href{http://xxx.lanl.gov/abs/physics/9812004}{
\mbox{\tt http://xxx.lanl.gov/abs/physics/9812004}}

\bibitem{grtensor}
{\em GRTensorII: Online documentation and information}.\hfill{}\\
Available at:~\href{http://grtensor.phy.queensu.ca/}{
\mbox{\tt http://grtensor.phy.queensu.ca/}}

\bibitem{lake2}
{\em GRTensorII demonstration page---general relativity \& geometry}.
Available~at:~\href{http://grtensor.phy.queensu.ca/NewDemo/demo.html#classic}{\mbox{
\tt http://grtensor.phy.queensu.ca/NewDemo/demo.html\#classic}}

\bibitem{redten}
J.~F.\ Harper and C.~C.\ Dyer:
{\em Tensor Algebra with REDTEN}.
Available 
at:~\href{http://www.scar.utoronto.ca/~harper/redten/root.html}{
\mbox{\tt http://www.scar.utoronto.ca/\~{}\,harper/redten/root.html}}

\bibitem{hartley}
D.~Hartley:
{\em Overview of computer algebra in relarivity} (1996),
in \cite{hehl}, pp.\ 173--191.

\bibitem{hehl}
F.\,W.\,\,Hehl, R.\,A.\,\,Puntigam, and H.\,\,Ruder, editors: 
{\em Relativity and
  Scientific Computing}, Springer-Verlag, Berlin, 1996. 

\bibitem{riccir}
J.~Kadlecsik:
{\em RICCIR: Ricci calculus package in REDUCE} (1996).
Available at:\hfill{}\\
\href{http://www.kfki.hu/cnc/szhkpub/riccir/riccir.html}{
\mbox {\tt
  http://www.kfki.hu/cnc/szhkpub/riccir/riccir.html}}

\bibitem{EinS}
S.~A.~Klioner:
{\em EinS: A Mathematica package for calculations with indexed
  objects. Information}. Available at: \hfill{}\\
\href{http://rcswww.urz.tu-dresden.de/~klioner/eins.html#abstract}{\mbox{\tt 
http://rcswww.urz.tu-dresden.de/\~{}\,klioner/eins.html{\#}abstract}}

\bibitem{krasinski}
A.~Krasi\'nski:
{\em The newest release of the ortocartan set of programs for algebraic
calculations in relativity}. Gen.~Rel.~Grav. {\bf 33} 145--161 (2001).

\bibitem{lake}
K.~Lake:
{\em GR 15 proceedings A5(ii) computer methods in GR: Algebraic
computing}.~Available~at:~\href{http://xxx.lanl.gov/abs/gr-qc/9803072}{\mbox{\tt 
http://xxx.lanl.gov/abs/gr-qc/9803072}}

\bibitem{ricci}
J.~M.~Lee:
{\em Ricci: A Mathematica package for doing tensor calculations in
  differential geometry. Documentation and source code}.\hfill{}\\
\href{http://www.math.washington.edu/~lee/Ricci/}{\mbox{
\tt http://www.math.washington.edu/$\sim$lee/Ricci/}}

\bibitem{sheep}
M.\,A.\,H.\,\,MacCallum: {\em Sheep: Information and source code}.\\
\href{http://www.maths.qmw.ac.uk/hyperspace/#ftp}{\mbox{
\tt http://www.maths.qmw.ac.uk/hyperspace/{\#}ftp}}

\bibitem{mac2}
M.\,A.\,H.\,\,MacCallum:
{\em Computer algebra and applications in relativity and gravity},
in A.~Macias, T.~Matos, O.~Obregon, and H.~Quevedo, editors, {\em
  Recent Developments in Gravitation and Mathematical Physics: Proceedings of
  the First Mexican School on Gravitation and Mathematical Physics}. World
  Scientific, Singapore, 1996.

\bibitem{mac1}
M.\,A.\,H.\,\,MacCallum and J.\,E.\,F.\,\,Skea:
{Sheep: A computer algebra system for general relativity} (1994),
in \cite{reboucas}, pp.\ 1--172.

\bibitem{DermottCA} J.\,D.\,\,McCrea, {\em REDUCE in General Relativity and
    Poincar\'e Gauge Theory} (1994), in \cite{reboucas}, pp.\ 173--263. 
See the
  library \href{ftp://ftp.maths.qmw.ac.uk/pub/grlib/}{\mbox{
{\tt ftp://ftp.maths.qmw.ac.uk/pub/grlib/}}}

\bibitem{macsyma}
Macsyma product information.
\href{http://www.macsyma.com/}{{\tt http://www.macsyma.com/}}

\bibitem{maple}
Maple product information.
\href{http://www.maplesoft.com/}{{\tt http://www.maplesoft.com/}}

\bibitem{riemann}
Maple package Riemann: Documentation and source code.\\
\href{http://www.astro.queensu.ca/~portugal/Riemann.html}{{\tt 
http://www.astro.queensu.ca/$\sim$portugal/Riemann.html}}

\bibitem{mathematica}
Mathematica product information.
\href{http://www.wri.com/}{{\tt http://www.wri.com/}}

\bibitem{cartan}
Mathematica package Cartan: Product information and excerpt of the Cartan
manual. Available at:\\
\href{http://www.universitetsforlaget.no/books/en/cartan/}{\mbox{\tt
http://www.universitetsforlaget.no/books/en/cartan/}}

\bibitem{pc}
L.~Parker and S.~M. Christensen:
{\em MathTensor: A System for Doing Tensor Analysis by Computer}. 
Addison-Wesley, Redwood City, 1994.

\bibitem{reboucas}
M.\,J.~Rebou\c{c}as, W.\,L.~Roque, editors,
{\em Algebraic Computing in General
    Relativity (Lecture Notes from the First Brazilian School on
    Computer Algebra, vol.\ 2)}. 
  Oxford University Press, Oxford, 1994. 

\bibitem{reduce}
Reduce online documentation and information.\\
\href{http://www.uni-koeln.de/REDUCE/index.html}{{\tt 
http://www.uni-koeln.de/REDUCE/index.html}}

\bibitem{schouten} J.\,A.~Schouten, {\it Tensor Analysis for
    Physicists.} 2nd ed.\ reprinted. Dover, Mineola, New York, 1989.

\bibitem{jimsk}
J.\,E.\,F.\,\,Skea et~al.:
{\em On--line invariant classification database}.\hfill{}\\
Available at: \href{http://www.astro.queensu.ca/~jimsk/}{
\mbox{\tt http://www.astro.queensu.ca/\~{}\,jimsk/}}

\bibitem{sheep2}
J.\,E.\,F.\,\,Skea:
{\em Applications of SHEEP} (1994).
Available at:
\href{http://www.can.nl/SystemsOverview/Special/Tensoranalysis/SHEEP/shpdrv.ps.gz}{
{\footnotesize\tt
http://www.can.nl/SystemsOverview/Special/Tensoranalysis/SHEEP/shpdrv.ps.gz}}

\bibitem{smh}
J.~Socorro, A.~Macias, and F.~W.~Hehl:
{\em Computer algebra in gravity: Reduce-Excalc programs for (non-)riemannian
  spacetimes. {\rm I}}, Comp.~Phys.~Comm. {\bf 115} 264-283 (1998). Also
available at: \hfill{}\\\href{http://xxx.lanl.gov/abs/gr-qc/9804068}{
\tt http://xxx.lanl.gov/abs/gr-qc/9804068}

\bibitem{sol}
H.~H.~Soleng:
{\em Tensors in Physics}.
Scandinavian University Press, Oslo, 1996.


\bibitem{ter}
S.~I.~Tertychniy:
{\em Searching for electrovac solutions to Einstein--Maxwell equations
  with the help of computer algebra system GRG$_{\rm EC}$} (1998).
Available at: \href{http://xxx.lanl.gov/abs/gr-qc/9810057}{\tt 
http://xxx.lanl.gov/abs/gr-qc/9810057}

\bibitem{tsantilis}
E.~Tsantilis, R.~A.~Puntigam, and F.~W.~Hehl:
{\em A quadratic curvature Lagrangian of Paw{\l}owski and R\c{a}czka : A
  finger exercise with Mathtensor} (1996),
in \cite{hehl}, pages 231--240.\\
Also available at: \href{http://arXiv.org/abs/gr-qc/9601002}{\tt
http://arXiv.org/abs/gr-qc/9601002}

\bibitem{deven}
A.~E.~M. van~de~Ven:
{\em Two-loop quantum gravity with the computer algebra program form} (1996),
in \cite{hehl} , pages 192--209.

\bibitem{wolf}
T.~Wolf:
{\em The program crack for solving pdes in general relativity} (1996),
in \cite{hehl}, pages 241--258.


\end{thebibliography}
\end{document}